# Circular photogalvanic effect in an inversion-symmetry-broken bilayer germanium nanosheet

**Taiki Nishijima [1], Ei Shigematsu[1], Ryo Ohshima[1], Keigo Matsushita[2], Akio Ohta[2,3], Masaaki Araidai[2], Junji Yuhara[2], Masashi Kurosawa[2], Masashi Shiraishi[1] and Yuichiro Ando[1,4,5]***

1. Department of Electronic Science and Engineering, Kyoto University, Kyoto, Kyoto 615-8510, Japan.

2. Graduate School of Engineering, Nagoya University, Nagoya 464-8603, Japan

3. Department of Applied Physics, Fukuoka University, Fukuoka 814-0180, Japan

4. PRESTO, Japan Science and Technology Agency, Honcho, Kawaguchi, Saitama 332-0012, Japan

5. Department of Physics and Electronics, Osaka Metropolitan University, 1-1 Gakuencho, Sakai, Osaka 599-8531, Japan

* Corresponding authors: Yuichiro Ando (yuichiro.ando@omu.ac.jp)

Abstract

Spin-to-charge conversion in monolayer and bilayer germanium(Ge) nanosheets was demonstrated via the circular photogalvanic effect (CPGE). The CPGE current generated in a spin-splitting state of the Ge nanosheet reached a maximum value when the thickness of the Ge nanosheet corresponded to bilayer germanene, indicating that the top layer of the bilayer Ge nanosheet mainly contributed to the spin-to-charge conversion. Because the hybridization of orbitals is suppressed by isolation from the bottom Al layer for the top Ge nanosheet, the observed spin-to-charge conversion has a possibility to be related to the intrinsic features of germanene with breaking of inversion symmetry.

The development of the fabrication process of single-layer graphene has become a catalyst for intriguing and exciting breakthroughs in condensed-matter physics, such as the experimental demonstration of ambipolar electrical transport by carriers, resembling massless Dirac fermions, and the creation of artificial van der Waals heterostructures[1–6]. Graphene remains a pivotal material for research in condensed-matter physics. However, it has the disadvantage of a zero band gap, which prevents efficient control of conductivity, thereby limiting the scope of engineering applications involving graphene. A negligibly small spin–orbit interaction (SOI) also limits phenomena related to spin-splitting states and topological quantum states.

Analogues of graphene consisting of other group-IV elements ("group-IV Xenes") in the periodic table, such as silicene (Si), germanene (Ge), stanine (Sn), and plumbene (Pb), are expected to show attractive characteristics originating from a considerable SOI and a buckled honeycomb structure[7–17]. Because of SOI, band inversion takes place at the K point in the reciprocal space for the materials, which is a feature of two-dimensional topological insulators under time-reversal symmetry[8,18]. The topological properties and band gap can also be controlled by an electric field perpendicular to the film plane, which is a decisive advantage over graphene[17,19–23]. These attractive properties have inspired vigorous synthetic research on group-IV Xenes, which have formed using various seed layers[12,24–40]. However, investigation of their characteristics has been strongly limited[41,42] because most seed layers exhibit metallic properties and thus act as stronger conductors than group-IV Xenes. The direct contact of group-IV Xenes and a metallic seed layer also induces modulation of the band structure around the Fermi level owing to orbital hybridization and prevents the intrinsic features of group-IV Xenes from being studied.

In this study, we focus on the SOI of a group-IV Xene. The magnitude of the SOI is approximately proportional to the fourth power of the atomic number ($Z^4$) in the simple hydrogenic model, and to the squared of $Z$ ($Z^2$) in the Landau-Lifshitz model, and group-IV Xenes have considerable SOI[43,44]. Therefore, spin-related phenomena originating from group-IV Xenes are expected to be dominant if the SOI of the seed layer is negligibly small. Here, we investigated spin-to-charge conversion in a Ge nanosheet on an aluminum (Al) layer fabricated by the segregation method. Al is well known to have negligible SOI and is one order of magnitude smaller than that of Ge[44–46]. Scanning tunneling microscope (STM) observation previously revealed that Ge atoms were arranged in a honeycomb structure after segregation[24], indicating the fabrication of germanene. In the present study, samples were fabricated in the same manner with a native aluminum oxide ($Al_2O_3$) layer to prevent oxidization even after air exposure[47]. Although we were unable to confirm the honeycomb structure of the samples in this study by STM because of presence of the $Al_2O_3$ layer, TEM observation revealed that a very thin Ge layer formed between the surface $Al_2O_3$ layer and the Al seed layer. Therefore, we use the term "Ge nanosheet" hereafter instead of germanene. It would also be expected that distortion of the band structure of the Ge nanosheet and $Al_2O_3$ thin layer would be suppressed around the Fermi level because of the large band gap of Al$_2$O$_3$. The inversion symmetry in our samples along the normal direction was inherently broken because the adjacent materials of the top and bottom interface of the Ge nanosheet were Al$_2$O$_3$ and Al, respectively. In this situation, Rashba-type spin splitting is expected, the efficiency of which depends on asymmetry of the wave function along surface normal, but still correlates positively with the strength of the SOI[48–51]. We detected the injection of spins into the Ge nanosheet by irradiation of circular polarized light and a helicity-dependent photocurrent that was caused by the spin-to-charge conversion, indicating a spin-splitting state of the Ge nanosheet. Such an optical approach also allows the highly sensitive detection of phenomena related to the Ge nanosheet, because the photocurrent generated in the Al layer is strongly suppressed by the reflection of light at the Al surface. The efficiency of spin-to-charge conversion reached a maximum value when the thickness of the Ge nanosheet corresponded to that of bilayer germanene, indicating the strong enhancement of the spin-to-charge conversion efficiency by electrical isolation of the Ge nanosheet from the Al seed layer.

A 25-nm thick Al(111) epitaxial layer on a Ge(111) substrate by evaporation through Joule heating underwent post-annealing in N$_2$ at atmospheric pressure for diffusion of Ge atoms into Al. Annealing temperatures ($T_{PA}$) of 200, 300, and 400 °C were used, and the annealing time was 60 min. The diffused Ge atoms were segregated between AlO$_x$ and Al layers during cooling of the sample because of the solid solubility limit of Ge atoms in Al crystals. The detailed procedure of sample fabrication and analyses of the crystal structure have been reported elsewhere[47]. Crystals of the segregated Ge layer showed a honeycomb structure, which was confirmed by STM, owing to the atomic arrangement of the epitaxial Al layer[24]. Although the sample for STM measurement was annealed after removal of the Al$_2$O$_3$ layer, we expect that all samples had a similar crystal structure because the Al$_2$O$_3$ layer does not have a clear crystal structure and would have had little impact on the atomic arrangement of Ge atoms.

Figure 1 shows the experimental setup for the helicity-dependent photocurrent measurement, where a photocurrent along the $y$-axis is collected. The contacts for photocurrent measurement were formed by Al wires that were bonded directly on the sample. The photocurrent was excited with a 642-nm semiconductor laser, the helicity of which was controlled using a linear polarizer and a quarter-wave plate (QWP). The light was periodically interrupted by an optical chopper with a frequency of 997 Hz, and the photocurrent was measured by a lock-in amplifier. The angle of incidence was defined with two parameters, the polar angle, $\theta$, and the azimuth angle, $\phi$. The irradiation position of light was precisely controlled with a three-axis piezo stage. Polarization dependence of the photocurrent was measured by rotating the QWP by the angle of $\psi$. At the output of the QWP, the light was linearly polarized at $\psi = 0°, 90°, 180°, 270°$ and right (left) circularly polarized at $\psi = 45°, 125°$ ($\psi = 135°, 315°$). To investigate the spin-related photocurrent, we measured the polarization dependence of the photocurrent. We expected spin-polarized electrons to be generated by the irradiation of circular polarized light. When spin-polarized electrons occupy spin-splitting states, a non-zero photocurrent is generated, which is known as the circular photogalvanic effect (CPGE)[52–56]. The CPGE current reverses its sign when the circular polarization is switched from right to left. For the CPGE measurements, samples were post-annealed at a temperature ($T_{PA}$) of 300 °C to achieve segregation of Ge unless otherwise indicated.

Figure 2(b) shows the photocurrent as a function of $\psi$ when the light was obliquely incident with $\theta = 35°$ in the $x$–$z$ plane ($\phi = 180°$) (Fig. 2(a)). The corresponding light polarization is shown on the horizontal axis. A clear reversal of the sign of the photocurrent was observed between the incidence of light with right and left circular polarization. This is analogous to the polarization dependence that has been observed in Rashba-type spin-splitting states[53,54,57] and the surface state of topological insulators[52,58–60].

The polarization-dependent photocurrent is phenomenologically expressed with a following equation[52]:

$$I_{\text{photo}} = C \sin 2(\psi + \psi_0) + L_1 \sin 4(\psi + \psi_0) + L_2 \cos 4(\psi + \psi_0) + D \tag{1}$$

$\psi_0$ is an offset angle based on the experimental setup ($\psi_0 \approx 4°$ in this study). The first term is a contribution from circular polarization with 180° rotation periodicity. The second and third terms stem from linear polarization with 90° rotation periodicity. The polarization-independent background, $D$, is generally attributed to the photovoltaic effect, Dember effect, and thermoelectric effect. However, in this experiment, it is mainly due to the thermoelectric effect because the position dependence of the irradiation is pronounced (see Supplementary Information, SI-I). By fitting this equation to the experimental results, we found that the circular-polarization-dependent term, $C \sin 2(\psi + \psi_0)$, which corresponds to the CPGE current had the largest contribution to the polarization-dependent photocurrent; this term is shown by a green line in Fig. 2(b). With the incidence in the $x$–$z$ plane, the in-plane projection of the angular momentum that circular polarization transfers to exited electron was polarized in the direction of the $x$-axis. This results in the nonequilibrium distribution of electrons along $k_y$ in the reciprocal space, as schematically shown in Fig. 2(c), where Rashba-type spin splitting may occur, and the CPGE current flows in the direction of the $y$-axis in real space.

When the angle of incidence was rotated in the $y$–$z$ plane, the CPGE current disappeared (Fig. 2(d) and 2(e)). That can be understood as the rotation of the aforementioned CPGE current-generation mechanism (Fig. 2(f)), whereby the CPGE current became undetectable with the experimental setup because it flowed in the direction of the $x$-axis. In the case of normal incidence, the CPGE current was not generated because the incident light only transferred out-of-plane angular momentum and the in-plane component was zero (Fig. 2(g), 2(h), and 2(i)). In addition, the symmetry of the germanene, $C_{3v}$, does not allow photocurrents induced by circular polarization[61]. These results are consistent with the CPGE current being caused by a Rashba-type spin-splitting state.

The CPGE current is phenomenologically expressed as[62]:

$$j_y = \sum_x \gamma_{yx} i(\boldsymbol{E} \times \boldsymbol{E}^*)_x \tag{2}$$

$$\mathrm{i}(\boldsymbol{E} \times \boldsymbol{E}^*)_x = \hat{e}_x E_0^2 P_{\mathrm{circ}} \tag{3}$$

where $j_y$ is the CPGE current in the $y$-direction, $\gamma_{yx}$ is a second-order pseudo-tensor, $\boldsymbol{E}$ is the electric field of the light, $\hat{e}_x$ is the $x$ component of the unit vector pointing in the direction of light propagation, $E_0^2$ is the amplitude or $\boldsymbol{E}$, and $P_{\mathrm{circ}}$ is the degree of circular polarization, which follows $P_{\mathrm{circ}} = \sin 2\psi$. $P_{\mathrm{circ}}$ is +1 (−1) for right (left) circular polarization. The CPGE current here is given by

$$j_y = \gamma_{yx} \hat{e}_x E_0^2 P_{\mathrm{circ}}, \tag{5}$$

where

$$\hat{e}_x = t_p t_s \sin\theta'. \tag{6}$$

Here, $\sin\theta' = \sin\theta / n$, $n$ is the refractive index, and $t_\mathrm{p}$ and $t_\mathrm{s}$ are the transmission coefficients for p and s polarizations, which are expressed with Fresnel's formula:

$$t_p t_s = \frac{4\cos^2\theta}{\left(\cos\theta + \sqrt{n^2 - \sin^2\theta}\,\right)\left(n^2\cos\theta + \sqrt{n^2 - \sin\theta}\right)}. \tag{7}$$

Thus, the CPGE current can be expressed by the following equation:

$$I_{\mathrm{CPGE}} \propto \frac{E_0^2 \sin\theta \cos^2\theta}{\left(\cos\theta + \sqrt{n^2 - \sin^2\theta}\,\right)\left(n^2\cos\theta + \sqrt{n^2 - \sin^2\theta}\right)}. \tag{8}$$

This equation gives a complex dependence of the CPGE current on $\theta$. The observed dependence of the CPGE current on $\theta$ in our study is well described by this formula (Fig. 3(a)). The experimental data and fitting curves both showed a peak around $\theta = 35°$. From Eq. (8), it is also expected that the CPGE current is reversed when the sign of $\theta$ is switched, which is related to the reversal of the spin direction (See S.I. Fig. S3). Furthermore, the CPGE current was proportional to the square of the incident electric field (i.e., the optical power), whereas the power dependence of the CPGE current shown in Fig. 3(b) showed a linear trend. The above analysis shows that our results agreed well with the phenomenological expression of the CPGE current, which is evidence that the observed circular polarization-dependent photocurrent stemmed from the CPGE induced by a Rashba-type spin-splitting state.

Although we confirmed that the CPGE current generation was due to spin-momentum locking in Rashba-type spin-splitting states, it is still not clear which layer of the samples generated the CPGE current. The contribution of the Ge substrate was expected to be negligible because most of the light was reflected at the Al surface and the intrusion light into the Al layer was attenuated before reaching the Ge substrate owing to the sufficiently thick Al layer (25 nm). Therefore, possible origins of the CPGE are the Ge nanosheet, Al layer, and/or their interfaces.

For the investigation of the CPGE-originating layer, CPGE currents in samples with different post-annealing temperatures, $T_{\mathrm{PA}}$, were compared. The amount of segregation of Ge atoms can be controlled by $T_{PA}$ because the solid solubility limit of Ge atoms in Al crystals increases with temperature in the eutectic system. After cooling the sample, the oversaturated Ge atoms segregated and formed a Ge nanosheet at the Al surface. The average thickness of the Ge nanosheet, $t_{Ge}$, on the Al layer was evaluated by X-ray photoelectron spectroscopy (XPS). Typical $Ge_{2p3/2}$ core-line spectra of the samples with various $T_{PA}$ are shown in Fig. 4(a). Clear Ge–Ge bonding peaks were observed for all samples and Ge–O bonding was sufficiently small except for $T_{PA}$ = 400 °C, owing to the $Al_2O_3$ layer for protection from oxidization. These spectra originated from the Ge nanosheet, and the contribution of the Ge substrate was negligible because the escape depth of the photoelectrons was much smaller than the Al thickness. We note that the Ge–Ge peak was clearly recognized even for the without annealing sample (before annealing in Fig. 4(a)). This is because intermixing of Ge and Al atoms took place even during Al deposition owing to sample heating. We confirmed that the intensity ratio of the Ge–Ge to Al–Al peaks did not appreciably change by changing the photoelectron take-off angle, indicating that Ge–Ge signal originated from near the surface. In addition, TEM observation revealed that the Al layer was sufficiently flat and continuous such that Ge–Ge bonding of the Ge(111) substrate could not be detected. The average thickness of the Ge nanosheet was estimated from the intensity ratio of the Ge line and the Al line. The detailed estimation procedure is described in the Supporting Information (section IV). The dependence of $t_{Ge}$ on $T_{PA}$ is shown in Fig. 4(b). Typical thicknesses of monolayer and bilayer germanene are also shown in the figure. The thickness of bilayer germanene includes the interlayer distance of each layer in addition to the height of buckling of the honeycomb structure. $t_{Ge}$ monotonically increased with $T_{PA}$, and $t_{Ge}$ was close to the thickness of bilayer germanene at $T_{PA}$ = 300 °C.

The polarization dependence of the photocurrent was determined in samples without post-annealing and for those post-annealed at 200, 300, and 400 °C in the same configuration as Fig. 2(a). In all samples, polarization-dependent curves and finite CPGE currents were observed (Fig. 4(c) and 4(d)). We note that even for the without annealing sample, there was a non-negligible CPGE current. We considered this to be the effect of the slightly segregated Ge atoms during Al deposition, which was confirmed in Fig. 4(a) and 4(b). However, the amplitude of the CPGE current for the without annealing sample was the smallest among all measured samples. The CPGE current reached a maximum at $T_{PA}$ = 300 °C, indicating that the Ge nanosheet (the thickness of which was close to the that of bilayer germanene) exhibited the highest spin-to-charge conversion efficiency. At $T_{PA}$ = 400 °C, the amplitude decreased because of the oxidization of the Ge nanosheet, as reflected in the XPS spectra. The strong dependence of the CPGE current on $t_{Ge}$ and the rapid reduction of the CPGE current by oxidization of the Ge nanosheet provide evidence that the CPGE current observed in Fig. 2 originated from the Ge nanosheet, because the conditions of the Al layer and Ge nanosheet/Al interface are not strongly dependent on $T_{PA}$.

Finally, we discuss spin-to-charge conversion in the sample post-annealed at 300 °C. It has been reported[21] that whereas bilayer germanene shows conducting edge states because of the quantum-spin Hall state (as clearly observed by STM), monolayer germanene does not show it because of distortion of the atomic arrangement and orbital hybridization between germanene and the seed layer. Density functional theory (DFT) calculations have revealed complex orbital hybridization of germanene adjacent to Al, resulting in a complex band structure with considerable Fermi circles[24,63]. In such a case, competing Fermi circles exhibit opposite signs in spin-to-charge conversion, resulting in reduction in the spin-to-charge conversion efficiency. A similar situation is also expected for the first layer of bilayer germanene. By contrast, the second layer of bilayer germanene is free from the influence of the Al layer and exhibits a simple band structure originating from the buckled honeycomb structure. Therefore, a simple Rashba-type spin-splitting state, as schematically shown in Fig. 2I, is expected under inversion symmetry breaking perpendicular to the film plane. We observed the highest spin-to-charge conversion efficiency for the Ge nanosheet when its thickness corresponded to bilayer germanene, suggesting a similar situation with bilayer germanene. The second layer of the Ge nanosheet is also expected to have inversion-symmetry breaking perpendicular to the film plane, satisfying the formation of Rashba-type spin-splitting states. It is not clear whether the obtained CPGE is related to the characteristics of germanene because the atomic arrangement was not revealed. Nevertheless, the utility of CPGE for investigating the characteristics of 2D materials, such as group-IV Xenes formed on a metallic seed layer with small SOI, is clearly demonstrated in this study.

In summary, spin-to-charge conversion of monolayer and bilayer Ge nanosheets was investigated via CPGE. The polarity of the CPGE current was reversed by changing the circular polarization, indicating the existence of spin-splitting states, such as Rashba-type spin splitting. The amplitude of the CPGE current strongly depended on the thickness of Ge nanosheet, $t_{Ge}$, indicating that the dominant contribution of CPGE was the Ge nanosheet. The maximum CPGE current was obtained for the sample with $T_{PA}$ = 300 °C, the $t_{Ge}$ of which corresponded to that of bilayer germanene. This indicates that that the second layer of the bilayer Ge nanosheet, which was isolated from the metallic Al layer, was the key for achieving highly efficient spin-to-charge conversion. The findings of this study is that symmetry breaking can create new functionalities in 2D materials, and that these functionalities enable us to evaluate the properties of 2D materials that were previously difficult to verify. Our results clearly show the usefulness of CPGE measurements for investigating 2D materials with considerable SOI, even on a metallic seed layer.

AUTHOR INFORMATION

**Corresponding Author**

**Yuichiro Ando** - Department of Electronic Science and Engineering, Kyoto University, Kyoto, Kyoto 615-8510, Japan, PRESTO, Japan Science and Technology Agency, Honcho, Kawaguchi, Saitama 332-0012, Japan, Department of Physics and Electronics, Osaka Metroplitan University, 1-1 Gakuencho, Sakai, Osaka 599-8531, Japan, Email: yuichiro.ando@omu.ac.jp

**Author Contributions**

T.N. conceived and conducted CPGE experiments supervised by M.S., Y.A., R.O. and E.S. K.M. and A.O. grew the sample and conducted XPS measurements. T.N., M.A., J.Y., M.K. and Y.A. analyzed the results. T.N. and Y.A. wrote the manuscript, with input from all authors. All authors discussed the results and reviewed the manuscript. The manuscript was written through contributions of all authors. All authors have given approval to the final version of the manuscript.

**Supporting Information**

The Supporting Information is available free of charge at XXXXXX.

Thermal-energy contribution to the photocurrent, 360° sweep data, Reversal of CPGE by switching the polar angle, and XPS analyses.

**Acknowledgments**

This work was partly supported by the Japan Society for the Promotion of Science (KAKENHI Grant Nos. 22H00214, 22H01524, and 20H02607), Japan Science and Technology Agency (JST), JST-PRESTO "Information Carrier" (Grant No. JPMJPR20B2), and MEXT Initiative to Establish Next-generation Novel Integrated Circuits Centers (X-nics, Tohoku University). We thank Adam Brotchie, PhD, from Edanz (https://jp.edanz.com/ac) for editing a draft of this manuscript.

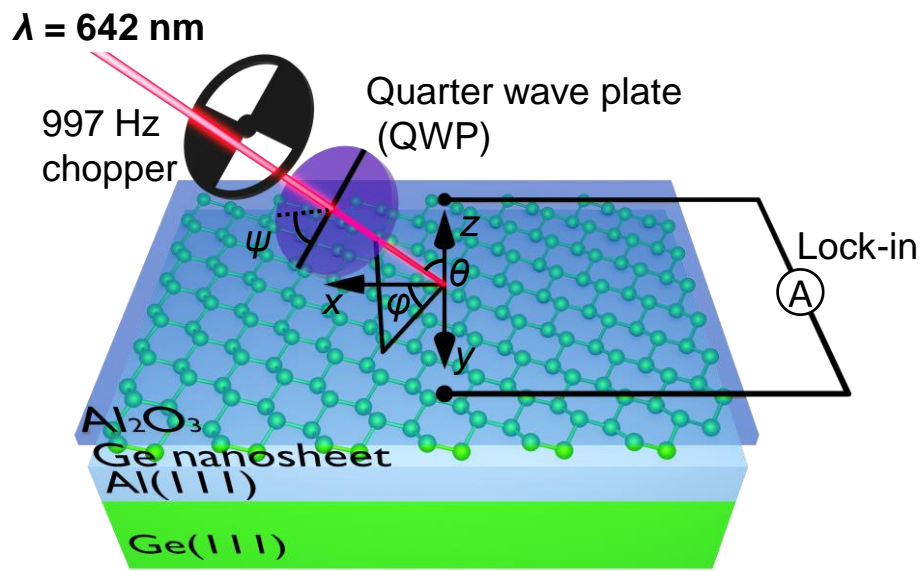

Fig. 1 T. Nishijima et al.,

**Fig. 1** Schematic illustration of the experimental setup for the measurement of the circular photogalvanic effect.

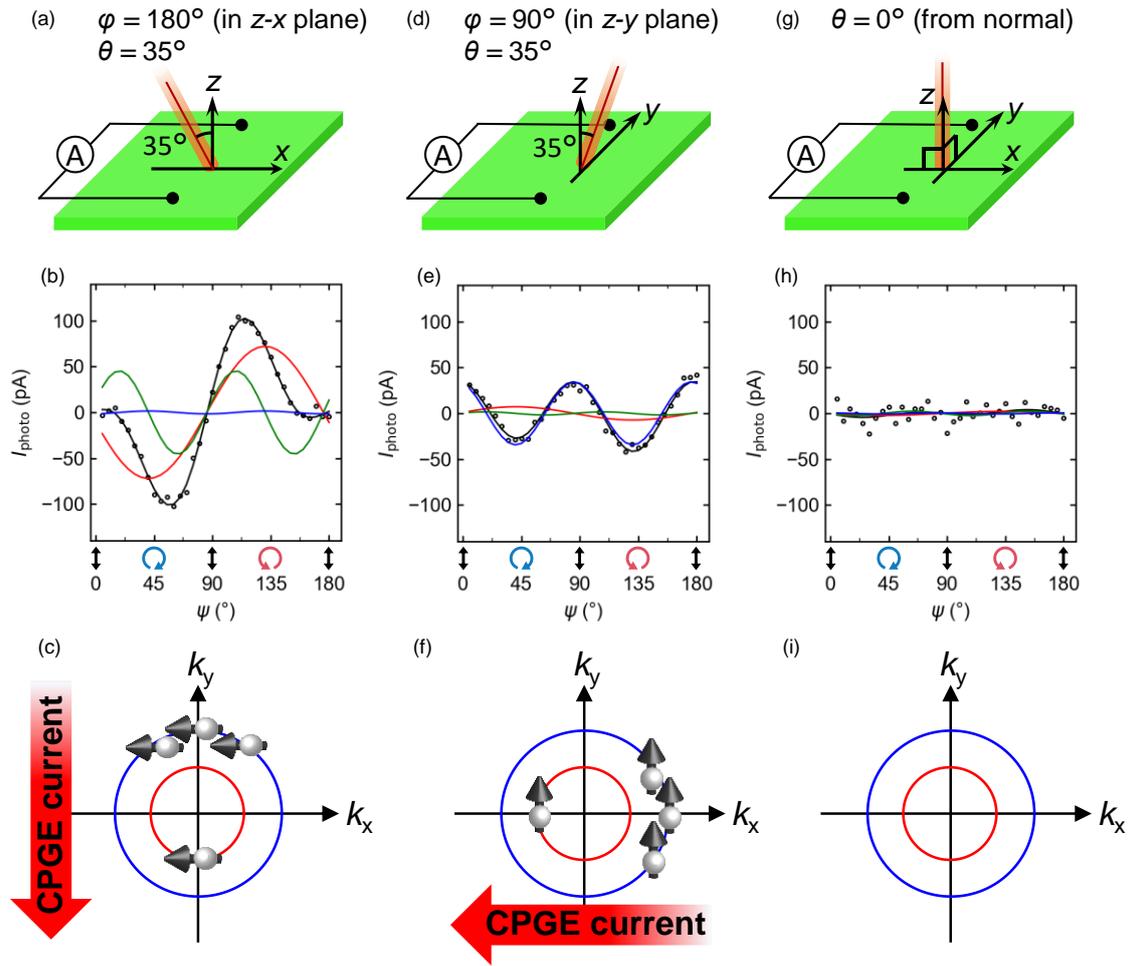

Fig. 2 T. Nishijima et al.,

**Fig. 2** Experimental setup, polarization-dependent photocurrent, and schematic representation of Fermi circles under spin accumulation. The light was irradiated in the *x*–*z* plane (a–c), in the *y*–*z* plane (d–f), and along the *z*-axis (g–i). The black solid lines in (b), (e), and (h) represent the fitting results with Eq. 1. The polarization-dependent terms obtained from the fitting results, $C \sin 2(\psi + \psi_0)$, $L_1 \sin 4(\psi + \psi_0)$, and $L_2 \cos 4(\psi + \psi_0)$, are shown by the red, green, and blue solid lines, respectively. Polarization of light controlled by the rotation of the quarter-wave plate is represented with arrows. The power of light was 4.8 mW.

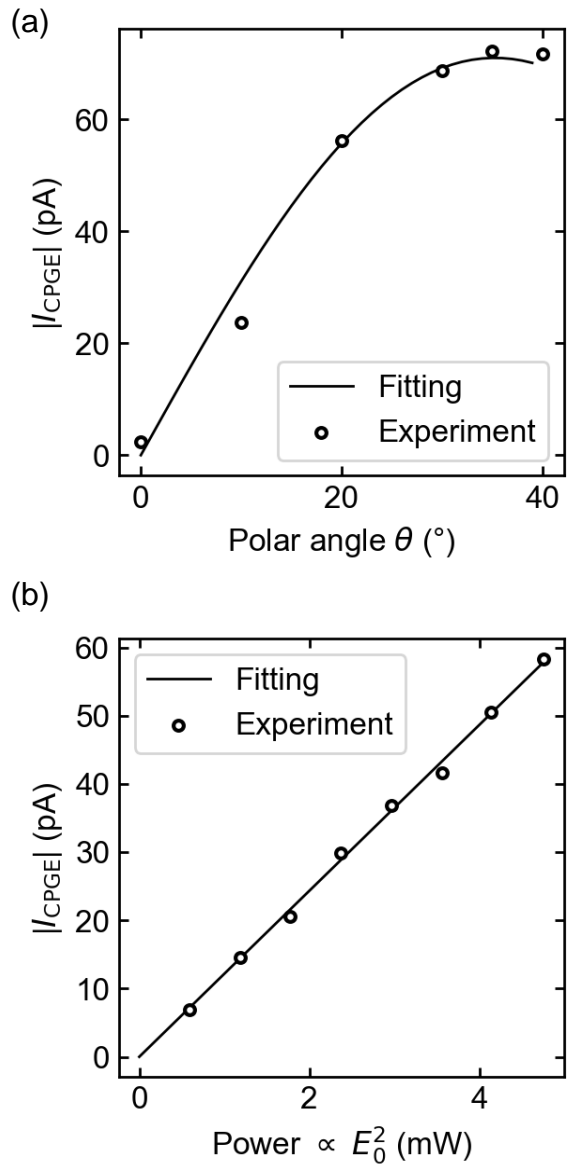

Fig. 3 T. Nishijima et al.,

**Fig. 3** (a) $\theta$-dependence of the CPGE current. The black solid line is the fitting curve with Eq. 8. The power of light was 4.8 mW. (b) Irradiation-power dependence of the CPGE current. The black solid line is a linear fit.

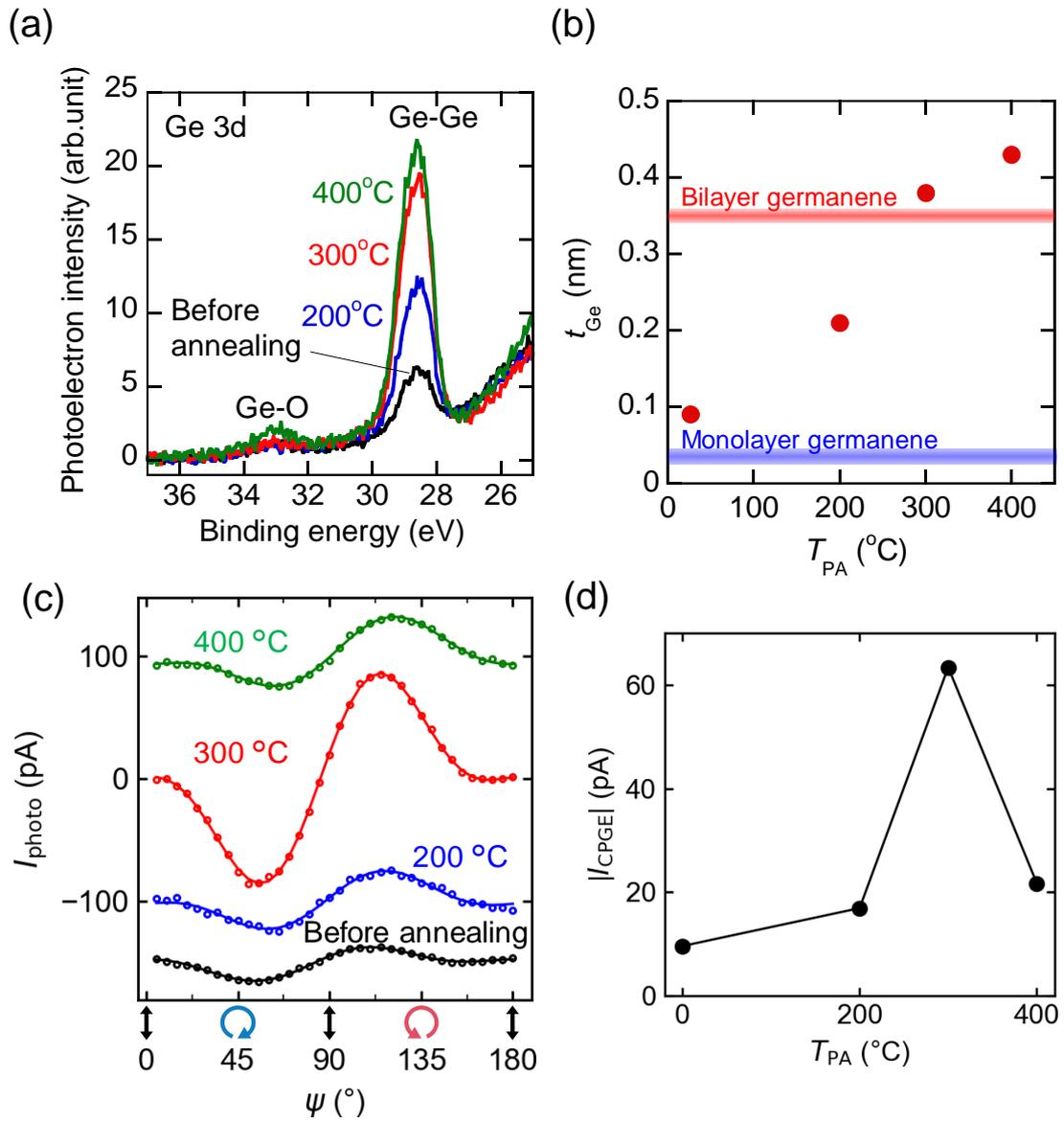

Fig. 4 T. Nishijima et al.,

**Fig. 4** (a) X-ray photoelectron spectroscopy of the Ge 3d core line for the samples before and after annealing at 200, 300, and 400 °C. (b) Dependence of the thickness of the Ge nanosheet, $t_{Ge}$, on the post-annealing temperature, $T_{PA}$. (c) Polarization-dependent photocurrent and (d) CPGE current for the samples before and after annealing at 200, 300, and 400 °C. The power of light was 4.8 mW.